\documentclass[12pt]{article}
\usepackage{graphicx,epsfig,amsmath,amssymb,color}

\begin{document}

\title{\vspace{-1.8in}
\begin{flushright} {\footnotesize LMU-ASC 69/12}  \end{flushright}
\vspace{0.3cm} {\ Universe Explosions }}
\author{\large Ram Brustein${}^{(1)}$,  Maximilian Schmidt-Sommerfeld${}^{(2,3)}$ \\
 \hspace{-1.5in} \vbox{
 \begin{flushleft}
  $^{\textrm{\normalsize
(1)\ Department of Physics, Ben-Gurion University,
    Beer-Sheva 84105, Israel}}$
 $^{\textrm{\normalsize
 (2) Arnold Sommerfeld Center fuer Theoretische Physik, LMU Muenchen,}}$ $^{\textrm{\normalsize\hspace{.2in} Theresienstrasse 37, 80333 Muenchen, Germany }}$
 $^{\textrm{\normalsize (3) Excellence Cluster Universe, Technische Universitaet Muenchen, }}$ $^{\textrm{\normalsize\hspace{.2in} Boltzmannstrasse 2, 85748 Garching, Germany }}$
 \\ \small \hspace{.3in}
    ramyb@bgu.ac.il,\ M.SchmidtSommerfeld@physik.uni-muenchen.de
\end{flushleft}
}}
\date{}
\maketitle

\begin{abstract}
A scenario for a quantum big crunch to big bang transition is proposed. We first clarify the similarities between this transition and the final stages of black hole evaporation. The black hole and the universe are thought of as quantum states. The importance of an external observer for understanding the big crunch to big bang transition is emphasized. Then, relying on the similarities between the black hole and the universe, we propose that the transition should be described as an explosion that connects the contracting phase to the expanding one.  The explosion occurs when entropy bounds are saturated, or equivalently when the states cease to be semiclassically (meta)stable.    We discuss our scenario in three examples: collapsing dust, a brane universe falling into a bulk black hole in anti-de Sitter space, and a contracting universe filled with a negative cosmological constant and a small amount of matter. We briefly discuss the late time observables  that may carry some information about the state of the universe before the transition.
\end{abstract}
\newpage

\section{Introduction}

The fate of a contracting universe heading towards a big crunch singularity has been studied for a long time \cite{rev1,rev2,rev3,rev4}. In the framework of general relativity the classical singularity theorems of Hawking and Penrose maintain that, generically, a contracting universe will reach a singularity within a finite time.  In some non-generic situations, such as closed empty de Sitter space, the universe contracts to a certain minimal size and then expands -- bounces -- following classical evolution.

It is widely expected that in a theory of quantum gravity the singularities will be smoothed; more precisely, one expects that a contracting universe will reach a state of high curvature and emerge from it in some other form. How this comes about and in what form the universe emerges from the high curvature epoch is a subject of discussion and controversy.

Many models of a possible big crunch -- big bang transition have been proposed; most, if not all of them, have some terms added to the Einstein equations such that the classical solutions of the modified equations exhibit the transition. In the framework of string theory this issue has been a subject of interest for decades. This was discussed in the context of  the pre-big-bang scenario \cite{pbb}, the so called Ekpyrotic universe \cite{ekpyrotic} and models involving null- or time-like singularities \cite{timelike} in string theory. More recently, the subject was investigated in the context of the gauge/gravity duality in  \cite{Hertog1,Hertog2,Hertog3} and \cite{Rabbarb1,Maldacena,Harlow,Rabbarb2,Rabbarb3}.

Rather than using the singularity theorems and energy conditions as tools to determine the fate of the universe, we will consider entropy bounds, which improve the diagnostic power of the classical singularity theorems, extending them to semiclassical situations \cite{ramyrev}.

An evaporating BH and a crunching universe will be argued to be subject to certain analogies, which will be employed to understand the destiny of a contracting universe. To be clear, most of our considerations/arguments refer to a single Hubble patch in a contracting universe. We think of the BH and the universe as semiclassical states. When they are large, treating them as semiclassical states makes sense since their decay widths (to be defined more precisely later) are much smaller than their masses. When they become smaller and reach a certain critical size, entropy bounds are saturated. We will interpret the saturation of the entropy bounds as signalling an instability of the BH or the universe. Below the critical size they can no longer be treated as (meta)stable semiclassical states. We will then argue that even when the BH and the universe are no longer semiclassically stable, the semiclassical estimate of their decay time is still approximately valid.

Our idea is to learn from the fate of an evaporating BH about the fate of a generic contracting universe. To decide on the fate of a contracting universe we need to observe it from an external point of view. For an observer living in the universe itself it is hard to keep track of the state of the universe, especially when it becomes unstable. The external point of view can be achieved either by observing a finite  contracting ``universe"  from outside, or by viewing the universe when it is embedded in a spacetime of different dimensionality.

When one observes the contracting universe from a higher dimensional point of view, the universe is a brane moving in a higher dimensional spacetime. Alternatively, one can view the universe from a lower dimensional point of view using the gauge/gravity duality \cite{hol1,hol2,hol3}. In this context, the universe is observed from its lower dimensional boundary. Both points of view turn out to be useful for understanding the fate of the universe.

We argue that the life of a contracting universe ends in an explosion, similar to the one in which a BH ends its life. The explosion occurs in all of space at approximately the same time and leads to a phase of expanding radiation dominated (RD) universe. The process that we envision is unitary, similar to the BH evaporation process, which is, as the modern view accepts, a unitary process. This means that the explosion connecting the big crunch to a big bang cannot generate entropy. Therefore, the total amount of entropy of the initial state and of the final state are equal, in contrast to many previous scenarios for which entropy production was a major obstacle for a bouncing or cyclic universe. Since the process is unitary and the total entropy is ``conserved",  the only way to get a large universe after the transition is to start with a large one from the beginning. If, for some reason, observations are made on a part of the universe after the transition has occurred, then this part can have a large entropy.

In our scenario, the big crunch -- big bang transition is intrinsically quantum mechanical and it is not very meaningful to think about it in terms of classical dynamics. The observables  that survive the transition are limited to  ``super-horizon" correlations imprinted during the contraction phase whose amplitude is frozen.

\section{Exploding black holes}

\subsection{Entropy bounds and their interpretation}

Bekenstein \cite{Bek1} proposed a bound on the entropy $S$ of a system whose energy is $E$ and whose linear size $R$ is larger than its gravitational radius, i.e. $R > 2 G E$,
\begin{equation}
\label{beb1}
 S\le ER.
\end{equation}
This is known as the Bekenstein entropy bound. Holography \cite{Boussorev} (see below) suggests that the entropy of any system is bounded by $ S_{HOL}\le A l_p^{-2}$, where $A$ is the area of the minimal space-like surface enclosing the system and $l_p$ is the Planck length.  The entropy bounds were further developed and versions applying to cosmological spacetimes were proposed \cite{Bek2,Causal,Boussorev}. For static spacetimes all entropy bounds are equivalent and imply that a BH is the most entropic state of matter.

Sometimes entropy bounds are interpreted as forbidding some state of matter or some properties of matter. Versions of this idea are referred to as ``The Species Problem" \cite{species1,species2,species3}. We argue that entropy bounds should be thought of as determining the region of parameter space in which BHs are semiclassical (meta)stable states.  From the point of view we are advocating the saturation of entropy bounds is a sign of an instability of BHs rather than signaling some fundamental restrictions on matter \cite{dvaliredi}.

We will restrict our discussions to theories containing a large number $N$ of weakly coupled light species. In such theories the effective Planck length is large, of order $\sqrt{N} l_p$, with $l_p$ the standard Planck length, which implies that the importance of certain quantum effects is enhanced. More precisely, under these conditions there are states for whose evolution some quantum (gravity) effects are crucial while the typical curvature is small, of order $1/N$ in Planck units. In the above discussion and in the following we ignore numerical constants since they are not important to the idea that we wish to present. Also, we discuss the entropy bounds in 4D. Re-inserting the numerical coefficients and extending the calculations to different dimensions is straightforward.

Let us consider quantum states in the part of the Hilbert space corresponding to a certain region of spacetime which is initially occupied by  a BH. The number of states in this part of the Hilbert space that are ``BH-like" is approximately $e^{S_{BH}}$   and the number of states that are ``radiation-like" is approximately $e^{S_{rad}}$. When $S_{BH}$ is larger than $S_{rad}$, a typical state is predominantly ``BH-like", but when $S_{BH}$ under time evolution becomes smaller than $S_{rad}$, the state becomes predominantly "radiation-like". For this evolution to be possible one needs that the time evolution can mix all the relevant states of the Hilbert space sufficiently rapidly.

To be specific, consider a black hole of mass $M$. Its size is given by its Schwarzschild radius $R_S=2GM$; its (Hawking) temperature is $T_H=1/R_S$. Let us compare the free energy of the BH, $E-TS$, to that of radiation consisting of $N$ species in thermal equilibrium at temperature $T=T_H$ in a region of space of size $R=R_S$. In fact, we will actually consider radiation in the same region of space that the BH occupies. The total energy of the radiation is $E_{\rm rad}\sim N T^4 R^3= N/R_S$ and its entropy is $S_{\rm rad}\sim N T^3 R^3=N$, while the energy of the BH is $E_{\rm BH}=M\sim R_S/l_p^2$ and its entropy is $S_{\rm BH}\sim R_S^2/l_p^2$.

So the free energy of the radiation is of order $N/R_S$ and the free energy of the BH is  of order $R_S/l_p^2$. If one now considers a very large black hole (whose free energy is larger than that of radiation occupying the same region of space), which evaporates and thereby becomes smaller, one sees that when the BH reaches a size $R_S\sim \sqrt{N} l_p$ the free energy of the radiation is equal to the free energy of the BH.

We interpret this point as the point in parameter space where the BH becomes unstable. We wish to support this interpretation by looking at the decay width of the BH. We make, following \cite{BDV}, two assumptions about the decay of semiclassical BHs. One assumption is that they emit as black bodies, so that
\begin{equation}
\label{BB}
\frac{dM}{dt}= - N T_H^4 R_S^2 = -N/R_S^2.
\end{equation}
The second assumption, following \cite{species3,BDV}, is that we require semiclassical BHs to obey similar conditions as other semiclassical states. In particular, we assume that the fractional change of the BH's mass is small on both the thermal and the light crossing time scales. The first condition
leads to $-\frac{1}{M} \frac{dM}{dt}< T_H$, while the second one to
$-\frac{1}{M} \frac{dM}{dt}< 1/R_S$.

To see what the implications are, let us look at the second condition. When the inequality is saturated, $-\frac{1}{M} \frac{dM}{dt}= 1/R_S$, one finds, using Eq.~(\ref{BB}), that
\begin{equation}
N T_H^4 R_S^3=M .
\end{equation}
This implies
\begin{equation}
N l_p^2=R_S^2
\end{equation}
and
\begin{equation}
N M_p^2=M^2 ,
\end{equation}
so, according to what we outlined before, at this point the entropy bounds get saturated. Also, as can be seen from the above equations, the BH looses a large fraction of its mass within one light crossing time.

As we will now show, this result also agrees with the interpretation that at this point the decay width of the BH becomes comparable to its mass, $\Gamma/M=1$.

Let us consider an ordinary system with many excited states and assume that it is highly excited and goes through a decay cascade. At the point where the mass and the decay width of a state become comparable, the system can no longer be considered as metastable and/or semiclassical.

Let us treat the BH in analogy to such a system and define $\Gamma$ as the inverse of the time before the first transition  to a lower energy state \cite{BDV}. According to this definition of $\Gamma$ one has $dM/dt= -\Gamma T$ for any black body. So in our case, $\Gamma= N/R_S$, and $\Gamma/M=1$ implies
$N/R_S= R_S/l_p^2$ which means $R_S=\sqrt{N} l_p$.

\subsection{The thermal decay rate is a good estimate for the duration of the final stage in BH evaporation}

We would like to understand whether when the BH becomes small and unstable we can still estimate its decay rate, so that we can determine the total lifetime of the BH. Above we have estimated it using the geometry and the thermal properties of the BH.  We will now argue that the thermal decay rate provides a good estimate even in the region of parameter space where the geometry of the BH changes significantly on the time scale set by the decay time, so that such an estimate cannot be expected a priori to be valid. This region of parameter space is characterized by the changes in the mass and the temperature becoming too fast or, equivalently, the decay width evaluated from the geometry becoming comparable to the mass, i.e. $\Gamma/M \gtrsim 1$.

Our idea is to treat the BH as a large object that decays to many particles each carrying a small fraction of the total energy. Then we can estimate the decay width from the phase space volume of the decay products.  For exactly $n$ massless particles in the final state it is given by
\begin{equation}
V_n(E)=\int \prod_{i=1}^n\left[ d^4p_i\ \delta(p_i^2)\ \theta(p_i^0)\right] \delta^4\left(P-\sum_{i=1}^n p_i\right).
\end{equation}
This is the microcanonical distribution of massless particles. The energy dependence of $V_n$ can be estimated in a straightforward way,
\begin{eqnarray}
V_n(E) &=& C \int \prod_{i=1}^n\left[\frac{d^3p_i}{E_i}\right]\ \delta^4\left(P-\sum_{i=1}^n p_i\right) \cr
&=& \widetilde{C} E^{2n-4}.
\end{eqnarray}

If, as mentioned above, the decay is dominated by many-particle final states we can treat this expression in a statistical way and define the relevant microcanonical entropy, $S\equiv \log V_n$. Then, for large $n$
\begin{equation}
S = const. + 2n \log E.
\end{equation}
We can also define the microcanonical temperature by $1/T = dS/dE$, from which we obtain
\begin{equation}
\label{parttemp}
T = \frac{E}{2 n}.
\end{equation}
The reason that the temperature is $E/ 2n$ rather than the more common $E/ 3n$ is that we have assumed that all the decay products are massless.

When one allows the number of decay products to vary, then instead of the microcanonical ensemble one gets the grandcanonical ensemble. The average number of particles is then given by the standard expression
\begin{equation}
\label{partnum}
\langle n \rangle = E/ 2 T.
\end{equation}
Again, $2T$ rather than $3T$ because we are considering massless decay products.

If the matrix elements governing the decay of the object are not highly peaked at some number of decay products $n$ or, put in different terms, if the effective chemical potential that the matrix elements induce is not peaked,  then we know that the ensemble of the decay products will approximately have the characteristics of black body radiation with temperature equal to the temperature in Eq.~(\ref{parttemp}) and number of particles roughly equal to the number in Eq.~(\ref{partnum}).

This picture of the final moments of a BH and its decay products is supported by the expected behaviour of BHs in string theory when they reach the ``correspondence line" \cite{correspondence1,correspondence2}. The BHs become highly excited states in the perturbative spectrum. Such states decay mostly into soft massless particles. The decay rate is therefore approximately thermal, with the temperature determined by Eq.~(\ref{parttemp}).

\subsection{Explosion ends the evaporation of a BH}

\begin{figure}[t]
\vspace{-1.5in}
\hspace{-0.7in}\scalebox{.40}{\includegraphics{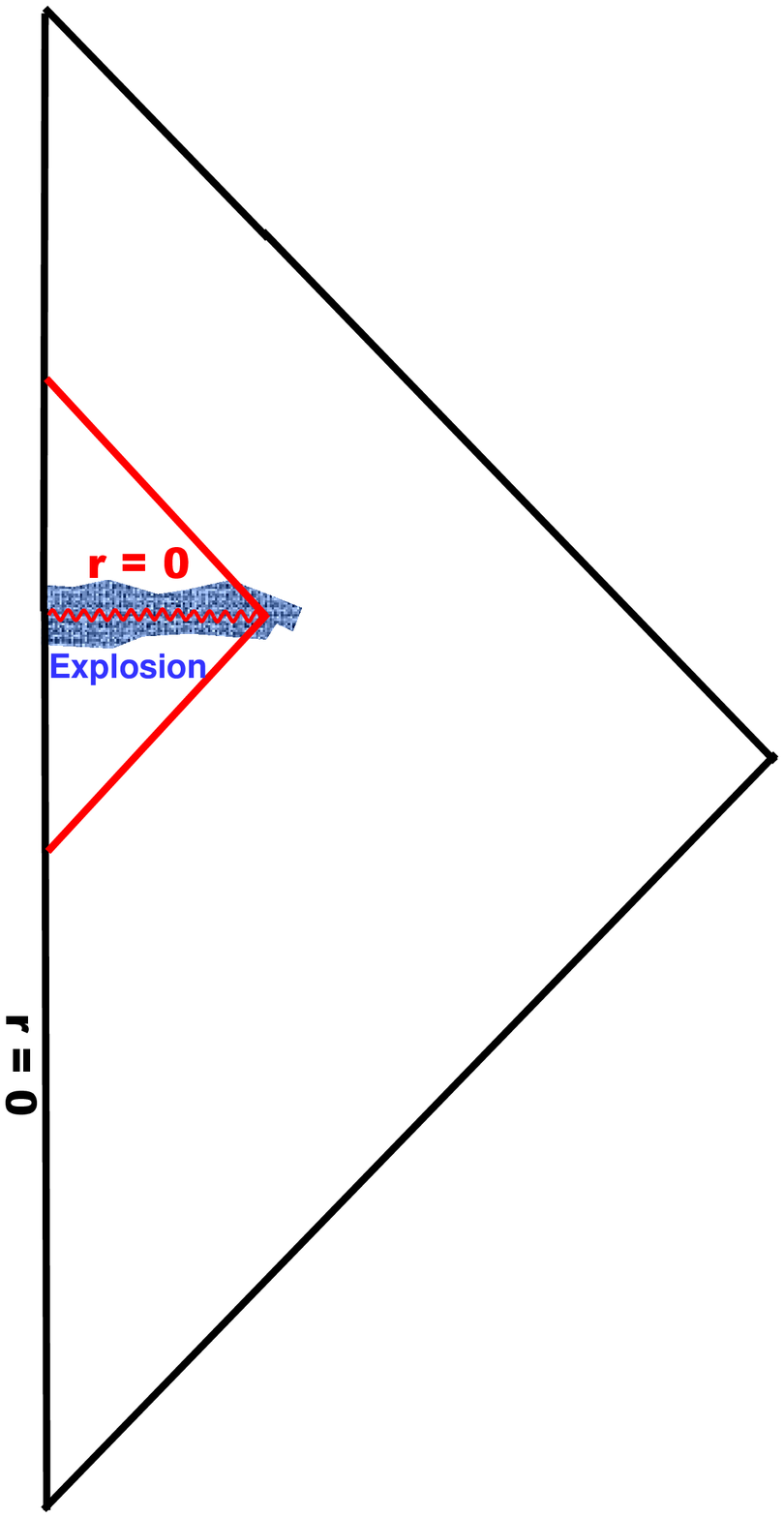}}
\hspace{-0.7in}
\scalebox{.40}{\includegraphics{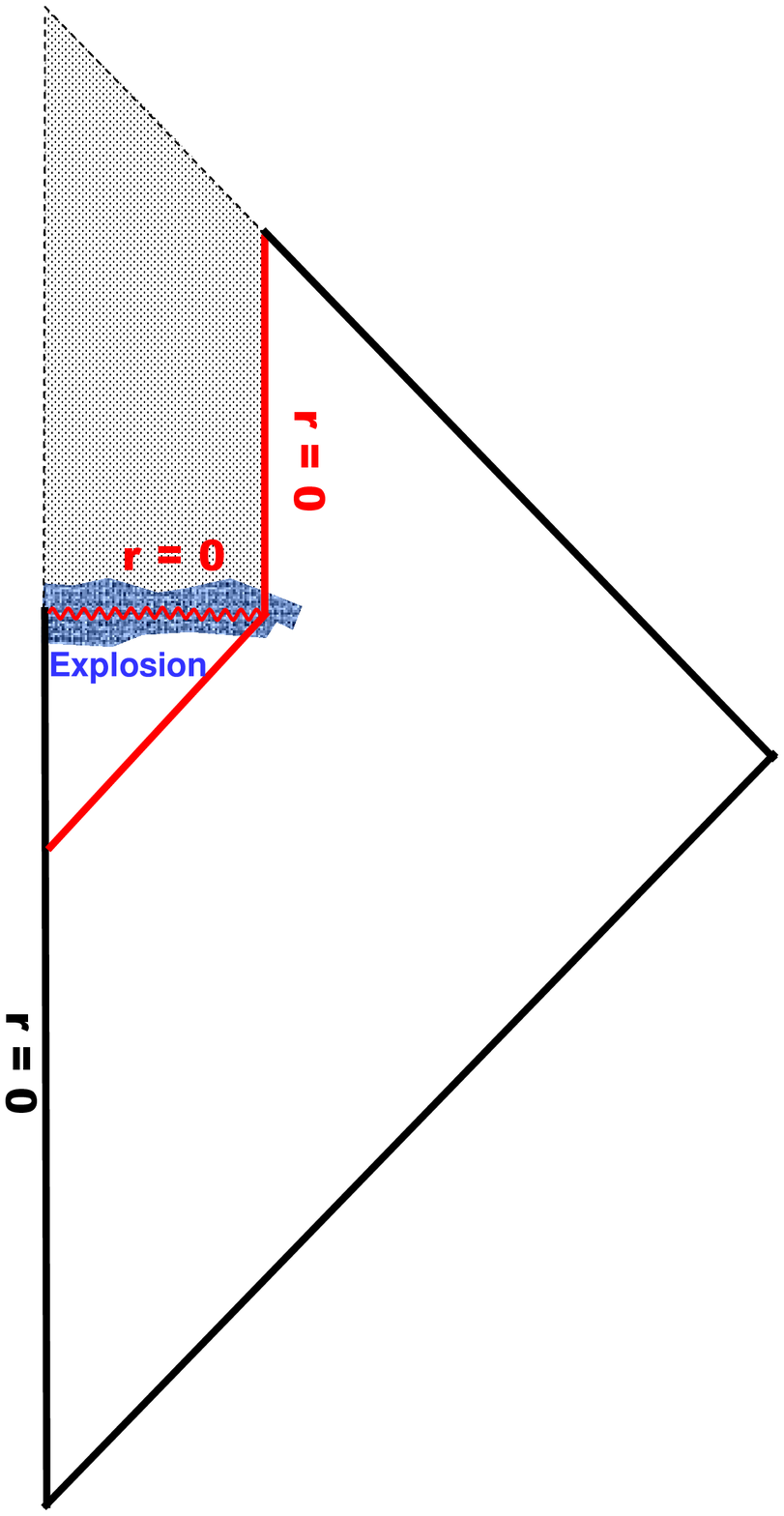}}
\caption{Two Penrose diagrams depicting the spacetime of an evaporating BH. The two diagrams depict essentially equivalent spacetimes. The right panel looks like the standard Penrose diagram of an evaporating BH while the left panel looks like a Minkowski space-time with an explosion region.  The volume of the excluded  shaded region on the right is Planckian and the distance between the various lines labeled $r=0$ is Planckian.}
\end{figure}

The conclusion of the above discussion is that even when the estimates based on semiclassical geometry are invalidated the decay still has the characteristics of a thermal decay. Since the temperature and other parameters are continuous when the BH ceases to be semiclassical, we can continue to use the thermal decay rate.

So we have found that when the BH reaches the point where $\Gamma/M=1$, which happens when its size is $R_S=\sqrt{N} l_p$, its lifetime is about $\tau= 1/\Gamma$, which in this case also means $\tau=R_S$. Our conclusion is that when the BH reaches the critical size, $R_S= \sqrt{N} l_p$, its energy gets converted into radiation in a time comparable to its light crossing time. This process is the most violent explosion possible: the whole energy in a certain region of space is converted into radiation in a time comparable to the time it takes light to cross the region. This is the shortest time allowed by causality for such a process to occur.

On the other hand, the total amount of energy released in the BH explosion should, in some sense, be thought of as small. The amount of energy released is of the order of one particle per light species in a region comparable to the particle quantum wavelength, so this state is not very different from the vacuum. It should be emphasized though that this state is also not very different from a BH. The reason is that the explosion has produced a state that is characterized by a certain amount of energy being contained in a region of size not much bigger than the Schwarzschild radius associated with that energy. So, in a sense, the state is also very close to a state that is "furthest" from the vacuum. We assume that one can nevertheless draw conclusions from the ``low density" property.

\section{Exploding Universe}
\subsection{Entropy bounds in cosmology}

Let us recall that cosmological spacetimes with horizons possess entropy. In this respect a universe is quite similar to a BH. The standard accepted example of this phenomenon is de Sitter space \cite{Gibbons}.  The argument is that for every observer in de Sitter space there is a region from which she cannot receive signals. The entropy is associated with the information contained in the inaccessible region. In general, a cosmological horizon, sometimes also referred to as a Hubble horizon, is not an event horizon, and what is outside it is not necessarily out of view for all times. Nevertheless, there are good reasons to associate an entropy with a cosmological horizon \cite{ramyrev,Boussorev}. The situation is clearest in the case that the cosmological horizon is approximately constant in size, which means that the universe behaves as a de Sitter space for a while.

We will assume that spacetimes with cosmological horizons possess entropy proportional to the area of the horizon in Planck units and discuss the fate of a contracting universe in this context.

Entropy bounds in cosmology were first discussed by Bekenstein \cite{Bek2}, who argued that if the entropy of the
visible part of the universe obeys the Bekenstein entropy bound, then the temperature is bounded and therefore certain cosmological singularities are avoided. Later, there have been many discussions following a similar logic.

The conclusion of the investigations of entropy bounds in cosmology was that cosmological spacetimes and BHs behave in a similar way. They saturate the entropy bounds when their size is ``Planckian".  For the universe this means that when the Hubble parameter reaches $ H= H_{MAX}\equiv\frac{M_p}{\sqrt{N}}$ entropy bounds are saturated. In an RD universe this is equivalent to the universe reaching a maximal temperature $T = T_{MAX}=\frac{M_p}{\sqrt{N}}$, as was anticipated by Bekenstein.

\subsection{Saturation of entropy bounds is a sign of instability}

As in the case of BHs and ordinary entropy bounds we will  argue that cosmological entropy bounds should not be thought of as constraining the  state of the matter whose energy-momentum tensor fuels the contraction. Rather they should be thought of as determining the region of parameter space in which a contracting universe can be considered as a semiclassical metastable state. From this point of view the saturation of entropy bounds is a sign of an instability of the contracting universe.

Following our discussion of BHs, let us compare the free energy of a universe, $E-TS$, to that of radiation consisting of $N$ species in thermal equilibrium at temperature $T=H$ in a region of space whose size is equal to the cosmological horizon radius, $R=1/H$. Then the total energy of the radiation is $E_{\rm rad}\sim N T^4 R^3=N H$ and $S_{\rm rad} \sim N T^3 R^3=N$, while the energy in a Hubble patch of the same size is $E_{\rm U}= \rho V\sim H^2 R^3/l_p^2$. So $E_{\rm U} \sim 1/(H l_p^2)$ and the entropy of the patch is $S_{\rm U}\sim R^2/l_p^2=1/(H^2 l_p^2)$.

We conclude that the free energy of the radiation is $N H$ and the free energy of the universe is  $1/(Hl_p^2)$. It follows that when a contracting universe, which was initially large, reaches a size $R=1/H=\sqrt{N} l_p$ the free energy of the universe becomes equal to the free energy of the radiation. We propose that, at this point, an explosion occurs in the sense that in the minimal time allowed by causality the universe is transformed into radiation.

\subsection{The necessity of an outside observer when determining the fate of a contracting universe}

Much of the confusion that prevailed in discussions on the fate of a contracting universe originates from viewing the evolution of the universe from within the universe. This is a natural choice for a viewing point since, one might argue, the universe is all there is, so its ``outside" is meaningless. But if an external ``objective" observer who can keep track of the fate of the universe could be found, she could record the evolution of the universe as it approaches the ``danger zone" and observe what happens subsequently.
It turns out that observers external to the universe are ubiquitous. All we have to do is to observe the universe from a spacetime of another dimensionality; either from a lower dimensional viewing point, e.g. from its boundary, or from a higher dimensional viewing point, in which case the universe is a brane in some ambient space being observed by a bulk observer. It is also possible to externally view a universe from a viewing point of the same dimensionality, however, then only the evolution of small universes can be recorded. In this case, the term ``universe" denotes only part of the state, as we will explain in detail below.

\subsection{Observables}

Observables at scales shorter than the horizon size at the time of the explosion are expected to be erased during the explosion due to the strong coupling and mixing between the different states. They will effectively thermalize. Of course, since the process is unitary no entropy is generated, so they are only approximately thermal. In any case, the relationship between the initial values of the short scale observables and their values after the explosion is very complicated. So, even if they can be calculated in principle, in practice this becomes impossible.

On the other hand, observables at large distance scales could carry some information about the state of the universe before the explosion.  For example, if the energy density of the Universe was slightly inhomogeneous, being different in different horizon size regions, then we expect that this information is expressed in the relative amplitudes and so in the properties of the emitted radiation. This issue came to the fore in previous investigations, for example those of the pre-big-bang scenario. The results on the behaviour of the large scale observables across the transition were found to be consistent with our expectations.

We believe that this is an interesting problem for further research.

\section{An exploding universe made from a spherical distribution of dust}

Let us consider the simplest model of a contracting universe. Of course, more sophisticated treatments exist in such a case, however, this model will serve as a good introduction to the concepts and issues that will be encountered in the more complex examples later on.

So consider a collection of (a finite number of) free non-relativistic massive particles which are distributed in a spherically symmetric way. The ensemble of particles extends up to a certain maximal radius $R_{tot}$ and its total mass is $M_{tot}$. Due to the mutual gravitational attraction of the particles the overall volume of the sphere that they fill out decreases and the density $\rho$, assumed to be homogeneous, increases. The standard derivation of the Friedmann equation from Newtonian gravity proceeds by considering spheres of radius smaller than $R_{tot}$, concentric to the large sphere of radius $R_{tot}$.  The equation of motion for a particle on such a sphere of radius $R$ reads
\begin{equation}
\frac{d^2R(t)}{dt^2} = -\frac{ G M(R(t)) }{R(t)^2} ,
\label{contract1}
\end{equation}
where $M(R(t)) = \int_0^{R(t)} 4\pi r^2 \rho(t) dr$. After multiplying this equation by $dR(t)/dt$ it can be integrated once. This is because the mass inside a comoving volume is conserved, i.e.
\begin{equation}
\frac{d}{dt} M(R(t))=0 .
\end{equation}
One obtains
\begin{equation}
\frac{1}{2} \left(\frac{dR(t)}{dt}\right)^2 -\frac{GM}{R(t)}=e.
\label{contract2}
\end{equation}
From this equation one infers that the integration constant $e$ is the total energy of a particle per unit mass. Its value is determined by the initial conditions. The initial conditions also determine the sign of $dR(t)/dt$ and thus whether the distribution of particles is expanding or contracting. We are interested in the latter case. To summarise, we are considering the case of pressureless spherical collapse.

If one now defines $H\equiv \dot{R}(t)/R(t)$, Eq.~(\ref{contract2}) multiplied by $2/R^2$ becomes the Friedmann equation
\begin{equation}
H^2 = \frac{2GM}{R^3}+\frac{2e}{R^2}.
\label{contract3}
\end{equation}
In the standard Friedmann equation, the sign of $e$ determines whether the the solution correspond to a closed, open or spatially flat universe.

We wish to consider a collection of particles with the property that every subset of the particles is always outside its Schwarzschild radius, so that an outside observer can comprehensively  describe their evolution. Let us recall here the necessity of an outside observer to keep track of and interpret the time evolution. For the concentric spheres of different radii that we are considering, the condition just stated reads $R(t)>R_S(R(t))$ for every $R(t)$. Equivalently,  $R(t)> 2G M(R(t))$ for every $R(t)$.

We are interested in a contracting distribution of particles, i.e. decreasing $R$, so we have to choose the initial conditions such that the subsequent evolution does indeed lead to a contraction. Since the first term on the RHS of Eq.~(\ref{contract3}) is proportional to $1/R^3$ and the second term is proportional to $1/R^2$, the first term will eventually dominate over the second term. We can choose initial conditions such that the first term becomes dominant while every subset of the particles is still outside their Schwarzschild radius. In this case we may,  for simplicity, set $e=0$.

So one ends up with the standard equation for a matter dominated, spatially flat, contracting FRW universe,
\begin{equation}
\frac {dR}{dt}=-\sqrt{\frac{R_S}{R}},
\end{equation}
whose solution is
\begin{equation}
R = \left(\frac{3}{2}\right)^{2/3} (R_S)^{1/3} (t_*-t)^{2/3}.
\end{equation}
Classically, the solution is singular.\footnote{Of course, before the singularity is reached the particles become relativistic and highly energetic so the assumption that they are free may well be invalidated.} However, as we have argued, when $H$ reaches the maximal critical value $1/(l_p \sqrt{N})$ an instability sets off. At this point the value of $R$ is
\begin{equation}
\label{minimalr}
R_{min}=\left(N l_p^2 R_S(R(t))\right)^{1/3}.
\end{equation}
This condition can be obeyed with $R_{min}> R_S(R(t))$ if $\sqrt{N} l_p > 2 G M(R(t))$, i.e. such that every subset of the particles is outside its Schwarzschild radius. The strongest condition comes from the whole distribution of particles, i.e. $\sqrt{N} l_p > 2 G M_{tot}$. Also, if $R_{min}> R_S$ then from Eq.(\ref{contract3}) we see that $dR/dt<1$, so the particles are still non-relativistic.

Here we are ignoring all other, non-gravitational, interactions of the particles. Their mass density at the minimal point is $M/R_{min}^3= 1/(N  l_p^4)$. So by making $N$ large we can keep the classical interactions small. We have assumed that we have $N$ weakly interacting  species, so if we choose the particles to be of these species, they can indeed remain weakly interacting also when quantum corrections are taken into account.

If $R_{min}> R_S$  it follows that the total mass of the system obeys,
\begin{equation}
\label{boundonmtot}
M_{tot} < \sqrt{N} M_p.
\end{equation}
Since the particles obey the Bekenstein entropy bound~(\ref{beb1}), it follows that $S < M_{tot} R_{min}$. Using Eq. \eqref{minimalr} for the whole set of particles, i.e. with $R_S=2G M_{tot}$, as well as Eq. \eqref{boundonmtot} one finds
\begin{equation}
S < N,
\end{equation}
as we have seen for BHs.  From a classical, geometric point of view, the universe could be considered as a large universe if $N$ is large. If we take the largest allowed value for $M_{tot}$,  $M_{tot} = \sqrt{N} M_p$,  then a volume of size $R_{min}^3 = N^{3/2} l_p^3$  could be quite large if one takes the fundamental cell volume to be $l_p^3$. However, the volume of the minimal phase space cell is $N$ times that, so from a quantum mechanical perspective, the universe that we have created is not the highly excited state that was perhaps expected. Since curvatures are always small, we do not expect the standard high curvature corrections to be important and therefore we can use the (semi)classical approximation when solving the equations of motion while the contraction rate is small enough.

\begin{figure}[t]
\vspace{-1.9in}\hspace{-1.2in}\scalebox{.45} {\includegraphics{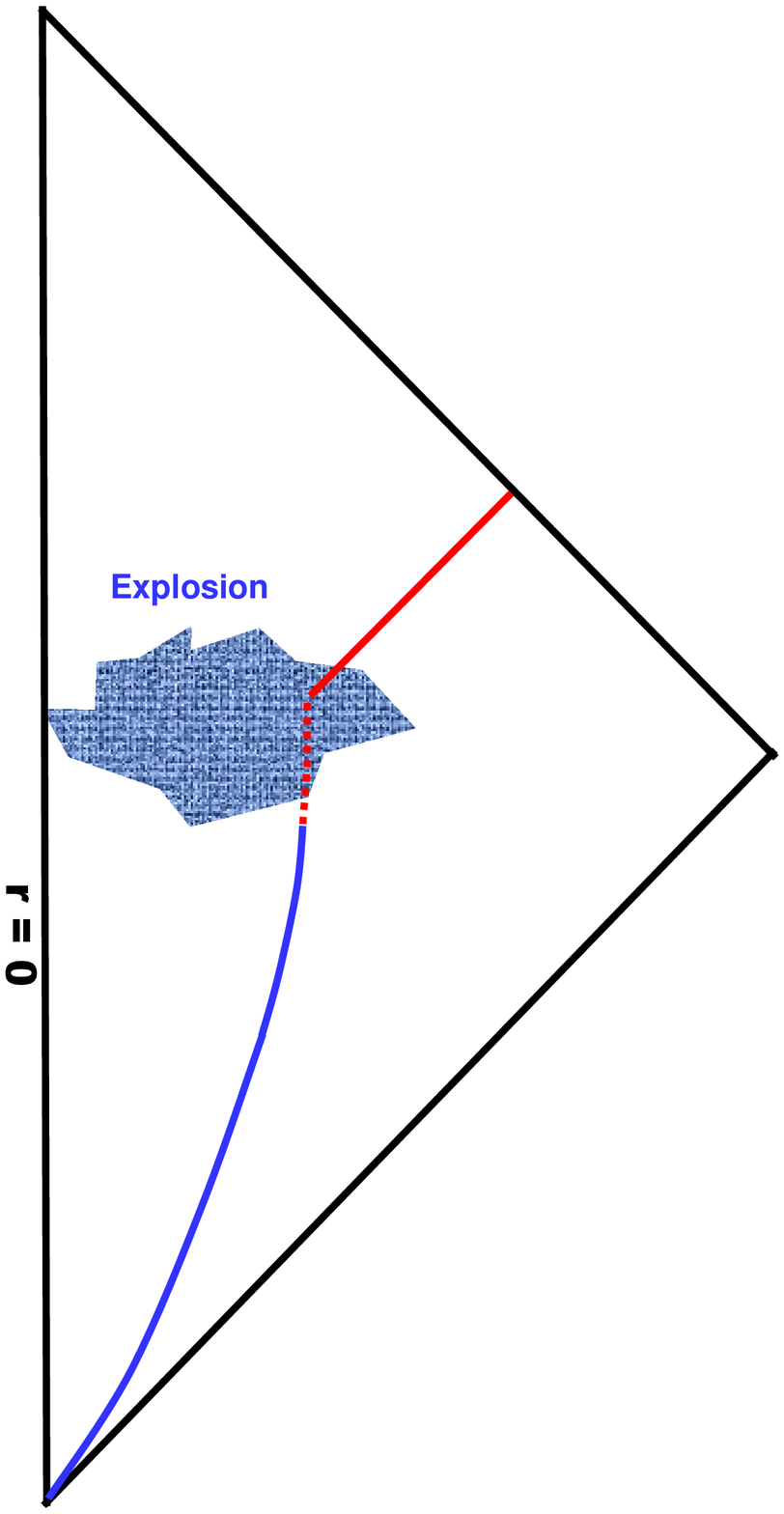}}
\hspace{-.7in}
\scalebox{.45} {\includegraphics{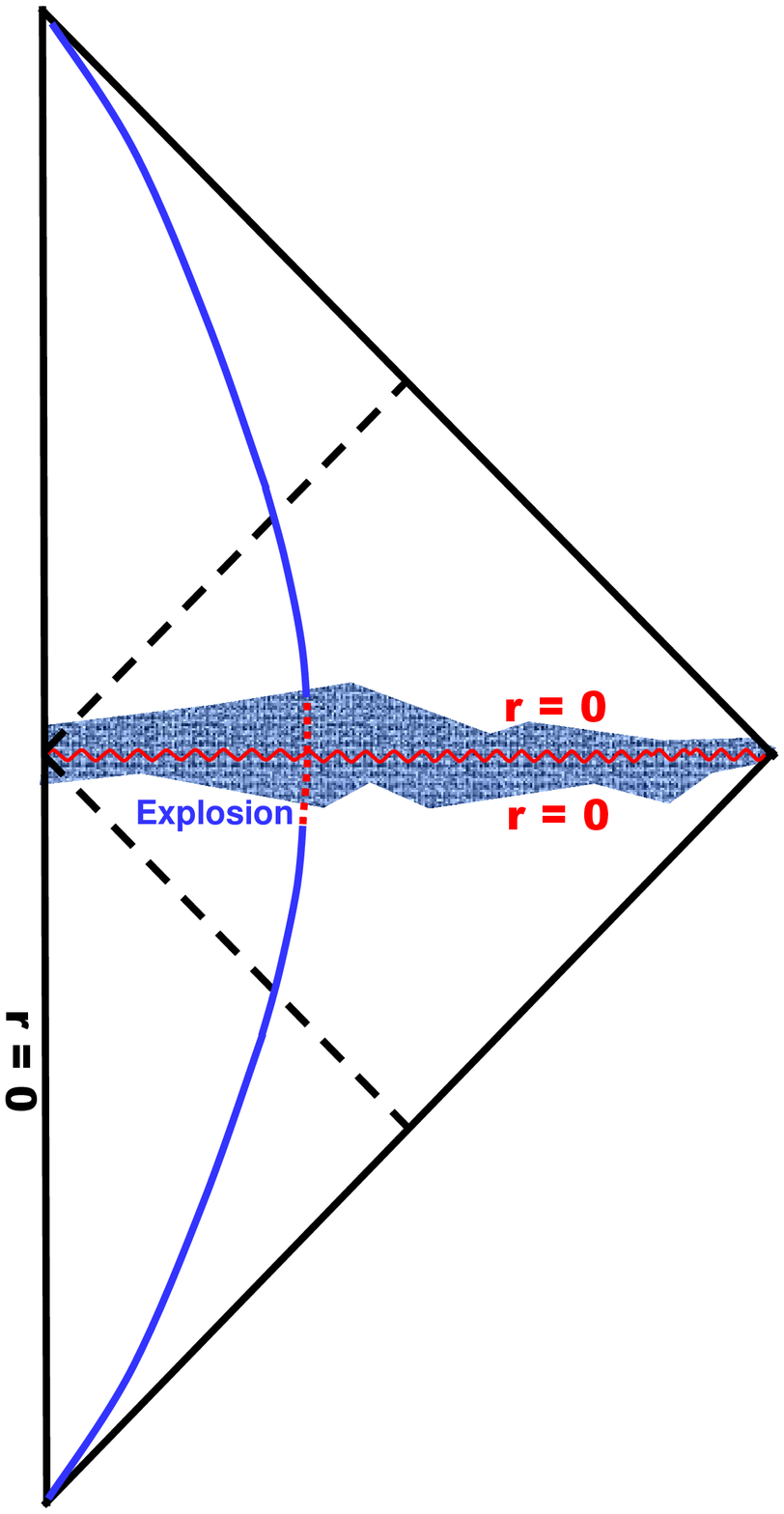}} \vspace{-.5in}
\caption{Penrose diagrams of an exploding Universe. Right: An insider view. A crunching FRW universe is joined through an explosion phase to another universe that starts with a bang. Left: An outsider view. A contracting universe explodes. If the explosion is instantaneous, an expanding shell of radiation is emitted. The case that the explosion is not instantaneous is discussed in the text.  }
\end{figure}

When the rate of contraction of the universe reaches the maximal critical value $H_{MAX}$, which corresponds to a minimal critical value for $R$, it becomes unstable. At this point, the free energy of radiation filling the same volume becomes larger than the free energy of a region of cosmological horizon size. An explosion occurs. If one approximates the explosion as instantaneous its outcome is quite simple:  some homogeneously distributed radiation with the same energy flies away isotropically to infinity. A contracting universe has been transformed into an expanding universe. When the explosion is not instantaneous, its detailed structure becomes important. This is an interesting problem for future research.

We have been able to describe a big crunch -- big bang transition of a small universe with minimal amount of entropy and expose the similarity of this process to the process of BH evaporation. Is it possible to describe in such a way a big crunch -- big bang transition of a large Universe? This is the subject of the next section.

\section{Exploding brane universe viewed from higher dimensions}
\subsection{An infalling FRW brane in AdS-Schwarzschild}

We can observe the explosion from a higher dimensional viewing point. The advantage is that we can describe the fate of a large universe. We accomplish this by throwing a 4D RD universe in the form of a brane moving in an AdS$_{5}$-Schwarzschild spacetime into the BH. Additional advantages are that the number of species on the brane has a clear geometric meaning and that the setup is well defined, at least in principle, by the boundary gauge theory.

The propagation of the brane in the AdS bulk  can be viewed in two alternative ways. One is the so called Randall-Sundrum  picture \cite{Randall:1999vf}. In this picture, a part of an AdS space is bounded by a brane on which a dual field theory ``lives". In the second picture, one thinks of  a complete AdS space, reaching up to its conformal boundary, with a probe brane moving on a geodesic in it. This is the so called mirage cosmology picture \cite{mirage1,mirage2}. In both cases the  brane evolution is identical to an FRW RD universe. Both descriptions will be used interchangeably in  the following discussion.

We will use the following representation for the
bulk spacetime,
\begin{eqnarray}
 \label{lnelem}
{\rm d}s^2=-H(R){\rm d}t^2+ \frac{1}{H(R)}{\rm d}R^2 + R^2{\rm
d}\Omega_3^2\, ,
\end{eqnarray}
where $H(R)=1+\frac{R^2}{L^2} - \frac{b^4 L^2}{R^2}$ vanishes at the BH horizon, $R_{S}$, and $b=\left(\frac{8 G_N^{(5)}}{3 \pi}\frac{M}{L^2}\right)^{1/4}$,
$M$ being the BH mass. $L$ is related to the cosmological constant of the AdS space and also to the brane tension $\lambda$, which is tuned in such a way as to make the effective cosmological constant on the brane vanish.

For the BH in AdS to be the dominant configuration over an AdS space with some thermal radiation as required for our analysis to be relevant, $b$ must be large,
$b\gg 1$ \cite{Witten}, that is, the black hole must be large and hot compared to the surrounding AdS$_{5}$. In this limit, the spatial curvature of the closed 4D universe is small, so it can be treated as flat. We can then write
\begin{equation}
R_S= b L,
\label{RSbrane}
\end{equation} and
\begin{equation}
T_H= \frac{b}{L},
\label{THbrane}
\end{equation}
where $T_H$ is the Hawking temperature of the BH. It should be pointed out that the AdS space that will be considered does not only contain a BH, but also radiation in thermal equilibrium with it.

The brane proper time is analogous to the proper time of a freely falling observer.  The evolution of the radial position of the brane
$R_{b}(\tau)$ is determined by an effective Friedmann equation:
\begin{eqnarray}
\label{brane}
\left(\frac{\dot{R_{b}}}{R_{b}}\right)^2=\frac{b^4 L^2}{R_{b}^4} -
\frac{1}{R_{b}^2}\, ,
\end{eqnarray}
where the dot means $\partial_\tau$.  The space curvature term is always negligible in the range that we are
interested, so we will ignore this term in the following. The first term on the RHS of the effective Friedmann equation \eqref{brane} looks like a contribution (to an ordinary Friedmann equation) of radiation. It is in this sense that the universe discussed here is radiation dominated. It should be pointed out though that the radiation is ``mirage radiation" rather than real radiation.

The number of species in the CFT is given
by $N = L^3/G_{N}^{(5)}$. The 4D and the 5D Newton's constants are related by $L G_{N}^{(4)}=G_{N}^{(5)}$ (again, we consistently ignore numerical factors).
As can be seen from the line element in Eq.~(\ref{lnelem}), the boundary of space
is a 3 sphere, so the CFT ``lives" on $S^3$. From Eq.~(\ref{brane}) one finds that the temperature measured on the brane is $T=b/R_{b}$, which is also in accordance with
the AdS/CFT correspondence. We notice that the boundary CFT temperature should not be confused with the Hawking temperature of the AdS BH as measured by a bulk observer located at $R$, which is given by $T_H /\sqrt{H(R)}$.

The entropy bounds are saturated when the temperature on the brane reaches its maximal value $T_{MAX}=M_p/ \sqrt{N}$. Expressing $M_p$ and $N$ in terms of 5D quantities, i.e. using $M_p=1/\sqrt{G_{N}^{(4)}}$ as well as the aforementioned relations between $G_{N}^{(5)}$, $G_{N}^{(4)}$, $L$ and $N$, one obtains $T_{MAX}=1/L$. Furthermore, as was just explained, there is a relation between the temperature on the brane and its position in the 5D space, namely $T R_b = b$. Thus, when the entropy bounds are saturated, $R_b = b/T_{MAX} = b L = R_S$. In other words, from the 5D point of view, the entropy bounds are saturated when the brane is about to fall through the horizon.

From Eq.(\ref{brane}) we see that at $R_b=R_S$, $\left(\frac{\dot{R_{b}}}{R_{b}}\right)^2 \approx 1/L^2$. This means that the size of the cosmological horizon on the brane, $\left(\frac{\dot{R_{b}}}{R_{b}}\right)^{-1}$, is  $L$, i.e. the AdS scale.

We now wish to estimate the energy density of the brane as it is about to fall into the BH \cite{tmax}. The cosmological constant on the brane is tuned such that its tension $\lambda$ exactly cancels the bulk cosmological constant, $G_{N}^{(5)} \lambda = 1/L$; this means that at horizon crossing, i.e. when $R=R_S$, the total energy of the brane  $E=\lambda R^3$ is given by
\begin{eqnarray}\label{enebra}
E =\frac{b^3 L^2}{G_N^{(5)}}\, .
\end{eqnarray}
Recall that $M\sim \frac{b^4 L^2 }{G_N^{(5)}} $, so it follows that $E/M\sim1/b$ which means that the total energy of the brane is much smaller than the BH mass, $E\ll M$.

\subsection{Absorption and emission of the FRW brane}

When the brane reaches the BH horizon, then from the point of view of the bulk it falls into the BH. According to our scenario, from its own point of view it explodes. We would like to show that the two points of view are actually consistent, thus supporting the idea that the explosion is the correct description on the brane.

When the brane falls into the BH, the BH gets a little larger and a little hotter. After some time the total energy of the infalling brane has been emitted and the BH returns to the original equilibrium state. The aftermath of this process is a shell of hotter Hawking radiation propagating towards the AdS boundary. This shock wave carries the entropy and energy of the  RD FRW universe that fell into the BH.  The only difference is that the radiation shell corresponds to an expanding universe rather than a contracting one.

So from the point of view of the bulk, the process of absorbing the brane in the BH and emitting the shell of Hawking radiation is equivalent to the explosion on the brane that ends the contraction phase and starts an expansion phase. Let us make these arguments more quantitative.

The effective emission rate of the BH is the difference of the total emission rate and the absorption rate. The absorption rate is equal to what the emission (and absorption) rate was before the brane fell into the black hole. When the brane falls into the BH, we assume that the mass of the BH changes instantaneously to $M+\delta M$ where $\delta M =E$ and $E$ is given by Eq.~(\ref{enebra}). Also, the temperature $T$ changes to $T+\delta T$ and the Schwarzschild radius changes to $R_S+\delta R_S$. The emission rate of the BH is
\begin{equation}
\label{emiss}
\frac{dM}{dt}= T^5 R_S^3,
\end{equation} thus,
\begin{eqnarray}
\frac{dM}{dt}|_{eff} = \frac{dM}{dt} |_{emit} - \frac{dM}{dt} |_{absorb} = (T+\delta T)^5 (R_S+\delta R_S)^3 - T^5 R_S^3.
\label{effemission}
\end{eqnarray}

Let us first present a rough estimate of the time it takes the BH to emit the brane. Using Eqs.~(\ref{THbrane}) and (\ref{RSbrane}) we can express the emission rate of the BH from Eq.~(\ref{emiss}) as
\begin{equation}
\frac{dM}{dt}= \frac{b^8}{L^2}.
\end{equation}
Since $M \sim b^4$ and $T \sim b$, $\delta T/T= 1/4\ \delta M/M$. Recall that the ratio $\delta M/M= 1/b$ is assumed to be  small. The effective emission rate is therefore given by (ignoring numerical factors),
\begin{equation}
\frac{dM}{dt}|_{eff}=  \frac{dM}{dt} \frac{\delta M}{M} = \frac{1}{b} \frac{dM}{dt}.
\end{equation}

Since the change in the emission rate is small, we can estimate the time $\Delta t$ it takes the BH to emit all the energy of the infalling brane and to come back to the equilibrium state by requiring that
\begin{equation}
\Delta t \ \frac{dM}{dt}|_{eff} = E  = \delta M,
\end{equation}
or equivalently,
\begin{equation}
\Delta t = \frac{E}{M}\frac{M}{\frac{dM}{dt}|_{eff}}.
\end{equation}

We may now substitute the expressions for $E/M$, $M$ and $dM/dt$ in terms of the bulk parameters and find that
\begin{equation}
\label{emissestimate1}
\Delta t = \frac{L^3}{G_N^{(5)}} \frac{1}{b^4} L.
\end{equation}

A more elaborate estimate of the time it takes for the emission of the shell of Hawking radiation can be obtained as follows. Using Eq.~(\ref{THbrane}) we can express $T$ as $T=(G_N^{(5)} M)^{1/4} L^{-3/2}$. Using Eq.~(\ref{RSbrane})  we can express $R_S$ as $R_S=(G_N^{(5)} M)^{1/4} L^{1/2}$. So we find for the emission rate $T^5 R_S^3=(G_N^{(5)} M)^{2} 1/L^6$. We can now substitute this expression into Eq.~(\ref{effemission}) and obtain (ignoring numerical factors),
\begin{equation}
  \frac{dM}{dt} |_{eff} = - \frac{(G_{N}^{(5)})^2}{L^6} M E.
\label{effemission2}
\end{equation}
Since the effective emission rate is also the rate at which the added energy $E$ is being emitted,
\begin{equation}
 \frac{1}{E} \frac{dE}{dt} = - \frac{(G_{N}^{(5)})^2}{L^6} M.
\label{effemission3}
\end{equation}
From Eq.~(\ref{effemission3}) we see that $E$ is emitted with a half-life of
\begin{equation}
  t_{1/2}= \frac{L^6}{M (G_N^{(5)})^2},
\end{equation}
or, expressing $M$ in terms of $G_N^{(5)}$, $L$ and $b$,
\begin{equation}
  t_{1/2} = \frac{L^3}{G_N^{(5)}}\frac{1}{b^4} L.
\label{emissestimate2}
\end{equation}
This confirms Eq.~(\ref{emissestimate1}).

\begin{figure}[t]
\vspace{-2.9in}\hspace{-.5in}\scalebox{.42} {\includegraphics{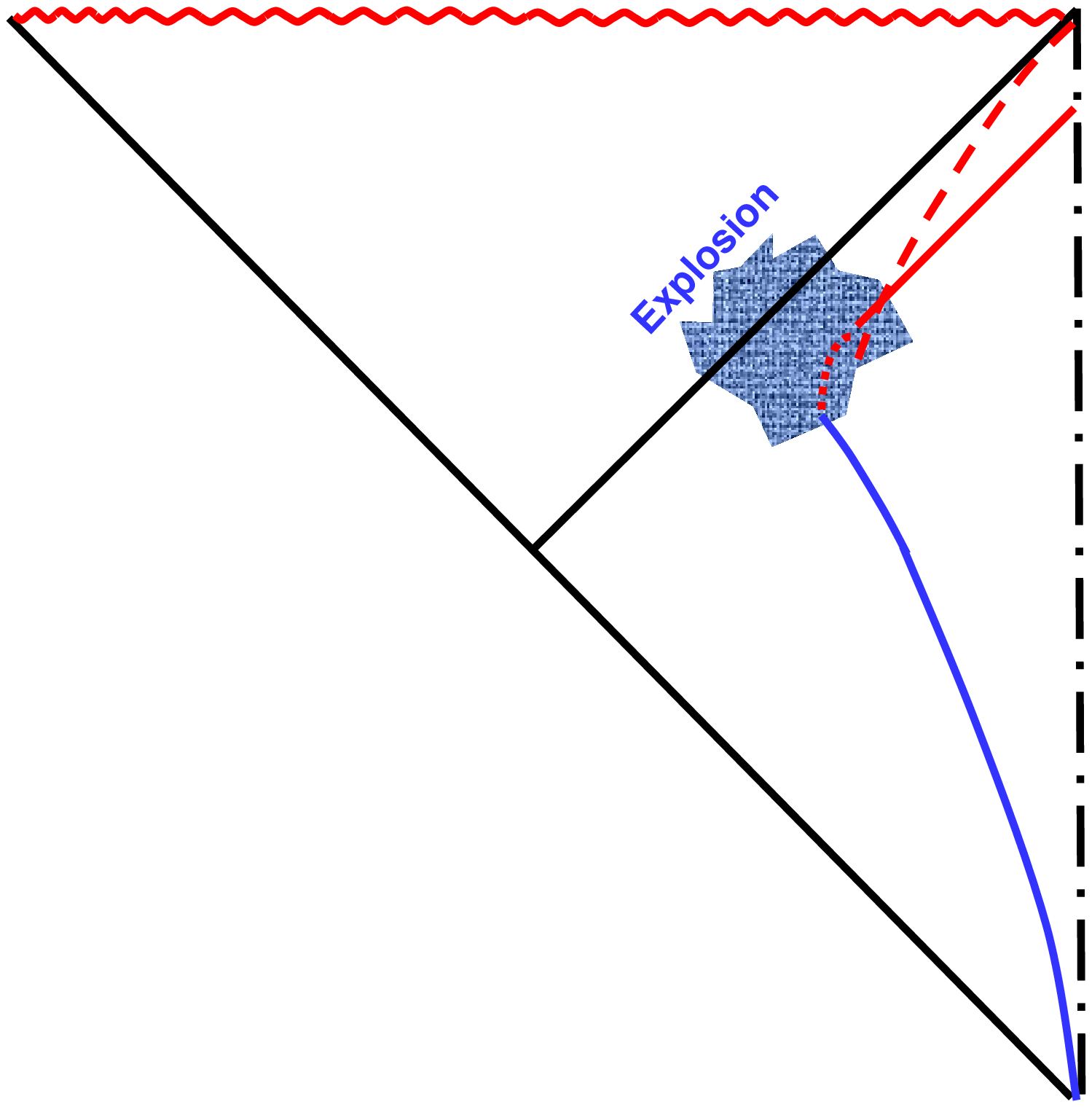}}\hspace{-0.6in}
\scalebox{.53} {\includegraphics{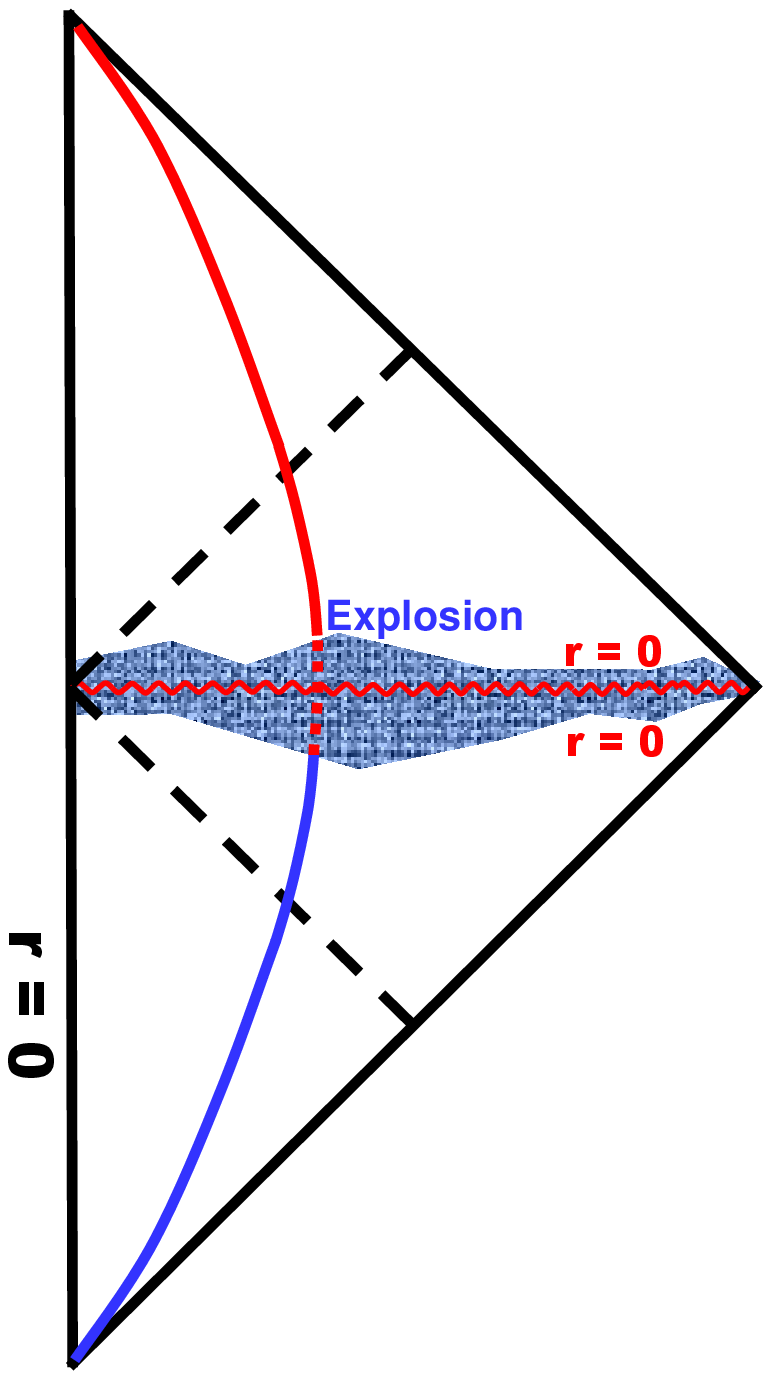}} \vspace{-.2in}
\caption{Shown in the left panel is a Penrose diagram of a brane universe falling into a BH (in blue) and the emitted shell of  radiation (in red) as viewed from the bulk.  The dotted red line describes an RD expanding universe. In the right panel the phases of a contracting, exploding and re-expanding universe as viewed from within the universe are shown.}
\end{figure}

The explosion time of the brane from the bulk point of view is $\Delta t_{explosion}= 1/H$ at the time that the brane reaches the horizon. As we have seen in Eq.~(\ref{brane}), at this point $1/H=L$. It follows that $\Delta t_{explosion}= L$.  So for a fixed ratio $L^3/G_N^{(5)}$ the ``reflection" time from the BH is shorter than the explosion time, meaning that the shape of the reflected brane is not distorted and its width is $L$, as it was before. The condition for this is
\begin{equation}
\label{reflectioncond}
 \frac{L^3}{G^{(5)}} \frac{1}{b^4} < 1.
\end{equation}
If the ``reflection" time scale is larger, then the emitted radiation shell is distorted with respect to the infalling one. In the extreme limit when the emission time is much larger we can approximate this by a BH simply becoming larger and heavier and reaching a new equilibrium point with its surrounding space. The study of the possible distortion effects is an interesting problem for further research.

In the approximation that the reflection is short and the explosion is instantaneous, the emitted shell is best described as a shock wave.  The shock wave carries the energy of the original infalling brane and also its entropy. This is because the BH remains in its original state and the entropy in the whole process must be non-decreasing. In the approximations that we are using, the entropy of the emitted shell is equal to the infalling one. So in this sense we may view the shock wave as a highly boosted RD universe. We expect that if the emission process is not instantaneous then the emitted shell will resemble an RD universe. This is an interesting problem for further research.

\subsection{The gauge theory interpretation}

If the brane is viewed as a probe brane then from the boundary  gauge theory perspective the brane represents
a homogeneous injection of energy into a single mode. From the microscopic point of view, this corresponds to singling out a brane from the stack of $N_c$ coincident branes and moving it a large distance away. One of the eigenvalues of the large matrix of excitations "moves towards the stack of branes." The coupling of this single brane to the rest of the branes is only through the very massive off-diagonal modes.

The motion of the brane towards the horizon represents a  change of the energy of the brane, still distributed in a homogeneous way, rather than a process of thermalization. Once the brane reaches the horizon, the average energy of the off-diagonal modes becomes comparable to the thermal energy. From a microscopic point of view, when the eigenvalue is large, its interactions through the off-diagonal heavy modes are suppressed, so it cannot thermalize. When it reaches the vicinity of the horizon, the off-diagonal modes become light enough and the coupling of the brane to its environment becomes strong.

From the bulk point of view, once the brane reaches the horizon, one expects it to be absorbed by the BH which becomes a little hotter and a little bigger.  This is consistent with the thermalization description above. However, if the reflection time is fast enough, the energy gets transferred back to some ``effective brane". This brane moves to the boundary and disappears. In this case, the gauge theory never thermalizes at a new temperature. The opposite limit in which the reflection time is slow, the BH mass increases and so does its temperature. From the gauge theory perspective this corresponds to thermalization.

Condition (\ref{reflectioncond})   implies for the gauge theory
\begin{equation}
\label{reflectioncond1}
 \frac{E}{M}< \frac{1}{\sqrt{N_c}}.
\end{equation}

This condition is probably related to requiring that the injected energy does not change the thermal state in a significant way and the interaction is still weak enough so the gauge theory does not have time to thermalize. A single brane has energy of order $1$ and the mass of the BH scales as $N_c^2$, so the condition on $E$ is expected to be satisfied.

\section{Exploding Universe viewed from lower dimensions}

We may also view the contracting universe from a lower dimensional point of view, using the AdS/CFT duality once more. We exploit the proposed duality between a contracting spatially curved FRW universe and a CFT on a codimension one de Sitter space. In this context, a contracting universe was studied previously by \cite{Hertog1,Hertog2,Hertog3,Rabbarb1,Maldacena,Harlow,Rabbarb2} with the purpose of understanding the fate of a crunching universe. The aforementioned proposal was elaborated in \cite{Rabbarb3}. The whole history of the universe, including a tunneling event from Minkowski into AdS, was considered. For us this whole history will not be relevant and hence we will not discuss it in detail.

If one considers an empty AdS space, i.e. the energy-momentum tensor contains only a cosmological constant term and no matter contributions, the contracting universe reaches a singularity in a finite time, however, this is a coordinate singularity. The universe reaches the singularity, re-expands and then contracts/expands cyclically ad infinitum. If one introduces only a little bit of matter into the FRW universe the coordinate singularity generically becomes, classically, a real space-like curvature singularity ``near" the coordinate singularity hypersurface.

Our scenario, when applied to this case, implies that when the imminent classical singularity is approached, quantum effects become strong, the contracting FRW universe explodes and turns into an expanding RD universe; the radiation then redshifts and the universe restarts the cycle.  This suggests that the cyclic evolution that one finds for the case without matter, when the singularity is a coordinate singularity, is also a good approximate description  in the more interesting case when, classically, the singularity appears to be a real physical singularity.

\subsection{Vacuum decay into Anti de Sitter space}

\begin{figure}[t]
\vspace{-1in}\hspace{+1.2in}\scalebox{.35} {\includegraphics{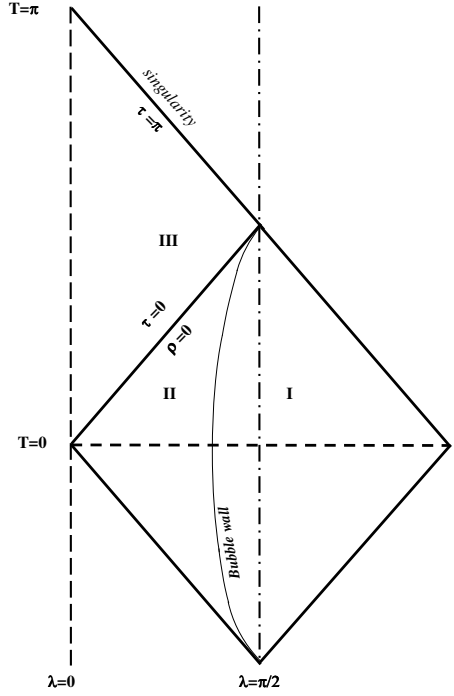}}
\caption{A Penrose diagram of the geometry described in the text. Region I is the Minkowski space (false vacuum), regions II and III are the AdS space; the metric in region III can be expressed in FRW form. In the case of an actual tunneling event from Minkowski to AdS, only the region above the dashed line $T=0$ exists.}
\label{fig:vacuumdecay}
\end{figure}

Let us first describe the geometry, a Penrose diagram of which is displayed in Fig.~\ref{fig:vacuumdecay}.

Regions I and II are those that Coleman and de Luccia obtained by analytically continuing their Euclidean solution describing vacuum decay. In this latter case,  only the part of the geometry above the horizontal dashed line $T=0$ exists. For concreteness,  the case of a decay from a state with vanishing cosmological constant to one with negative cosmological constant is considered.  However, we will not discuss the initial tunneling event in detail and we will consider the full geometry that is depicted in Fig.~\ref{fig:vacuumdecay}. The metric in regions I and II can be expressed as follows:
\begin{eqnarray}
ds_{\rm I}^2 &=& dr^2 + r^2 ( - dt^2 + \cosh^2(t) d\Omega^2 ), \quad r_0<r \quad 0<t
  \\
ds_{\rm II}^2 &=& R^2_{AdS} [ d\rho^2 + \sinh^2\rho ( - dt^2 + \cosh^2(t) d\Omega^2 ) ],
\quad 0 < \rho < \rho_0 \quad 0<t.
\nonumber
\end{eqnarray}
One can continue across the line $\rho=0$, i.e. the boundary between regions II and III in the above figure. ``Behind" that line, there is a further region of AdS space, in which the metric can be written in an FRW form:
\begin{eqnarray}
ds_{\rm III}^2 = R^2_{AdS} [ - d\tau^2 + \sin^2(\tau) ( d\chi^2 + \sinh^2(\chi) d\Omega^2 ) ],
\quad 0<\tau<\pi.
\label{ds2III}
\end{eqnarray}
A coordinate system that covers both regions II and III can be found. In this coordinate system the metric takes the form:
\begin{eqnarray}
ds_{\rm II+III}^2 = R^2_{AdS} [ - \cosh^2(R) dT^2 + dR^2 + \sinh^2(R) d\Omega^2 ],
\nonumber \\
\quad 0<T<\pi \quad 0<R \quad \cosh(R) \cos(T) < \cosh(\rho_0).
\label{ds2V}
\end{eqnarray}
The relation $\cosh(R) \cos(T) = \cosh(\rho_0)$ describes the location of the bubble wall, i.e. the boundary between regions I and II. The coordinates $R$, $T$ are related to the ones used above as follows:
\begin{eqnarray}
\cosh(\rho) = \cosh(R) \cos(T), \quad \tanh(t) = \frac{\sin(T)}{\tanh(R)},
\\
\cos(\tau) = \cosh(R) \cos(T), \quad \tanh(\chi) = \frac{\tanh(R)}{\sin(T)}.
\end{eqnarray}
These relations suggest a (formal) analytical continuation from one coordinate system to the other,
\begin{eqnarray}
\rho = i \tau \ , \ \chi = t + \frac{i \pi}{2} \ .
\end{eqnarray}
In order to draw the Penrose diagram in Fig.~(\ref{fig:vacuumdecay}), one defines a new radial coordinate via
\begin{eqnarray}
\cosh(R) = \frac{1}{\cos(\lambda)} \ .
\label{rlambda}
\end{eqnarray}
The conformal boundary of AdS space is located at $\lambda=\frac{\pi}{2}$.

The metric in region I and the metric in region II in the $\rho$, $t$ coordinates take the form of de Sitter slicings and  the geometry of the bubble wall is that of  de Sitter space.

Maldacena's proposal is that a dual description of the geometry just described is obtained by replacing its AdS part, i.e. regions II and III, by a conformal field theory living on the bubble wall. The dual description thus consists of a part of Minkowski space (region I) bounded by a codimension one de Sitter space, on which a conformal field theory lives.

So far, regions II and III, i.e. the interior of the bubble, have been treated as empty, spatially curved, AdS space. In this case, there is a coordinate singularity at the hypersurface $\tau=\pi$ where the FRW scale factor in Eq.~(\ref{ds2III}) vanishes. This hypersurface is described by $\lambda=\pi-T$ in the coordinates (\ref{ds2V}), (\ref{rlambda}) and is totally regular there.

\subsection{The crunching and exploding universe}

Upon introducing some matter, i.e. when considering a bubble interior that is not completely empty AdS space, the coordinate singularity at $\tau=\pi$ generically becomes a real curvature singularity ``near" this hypersurface. The geometry in this case is shown in Fig.~5.

Barbon and Rabinovici as well as Maldacena  proposed that the realistic case of a classical physical singularity can be described by deforming the dS field theory by an irrelevant operator. This operator becomes important for bulk observers that get close to the conformal boundary of AdS space. When the observer gets close enough to the conformal boundary, there will be a point when the effect of the irrelevant operator becomes so large that the evolution has to be changed in a significant way. At this point the dual field theory stops being semiclassical and one needs to invoke some new physics to be able to describe what happens.

Changes in the dS field theory that are homogeneous in the dS time correspond to setting initial conditions to the FRW evolution. We argue that from the region III FRW point of view this is equivalent to adding a source (some form of matter) that becomes important at later FRW times $\tau$. From a region II bulk observer  point of view, it is possible to interpret the state of a dS boundary with a large irrelevant operator as a highly excited state of the dS. We argue that even when the theory is not semiclassical,  this highly excited state is very similar to a thermal state.  If this similarity is correct then from the boundary point of view the irrelevant operator  describes a BH that is being created in the bulk (See \cite{Rabbarb2}). Our proposal is that  the explosion that marks the big crunch to big bang transition is viewed from the boundary as the final stages of evaporation of this BH. This is shown in Fig.~5.

Our proposal is consistent with Maldacena's proposal (and with the description of Barbon and Rabinovici) and provides a specific scenario for the fate of the universe when the irrelevant operator has become dominant. At this point a precise description in terms of the field theory is not available as it would require detailed knowledge of the cutoff scale physics. So we conclude that the explosion is dual to a state on the boundary that cannot be described by semiclassical physics in dS. In this sense, the explosion ``spills" to the AdS boundary.  It creates some highly excited state which lives in the UV on the whole AdS boundary, so in a way it is like a big bang from the point of view of the dual dS field theory.

\begin{figure}[t]
\vspace{-1.0in}\hspace{-0.25in}\scalebox{.30} {\includegraphics{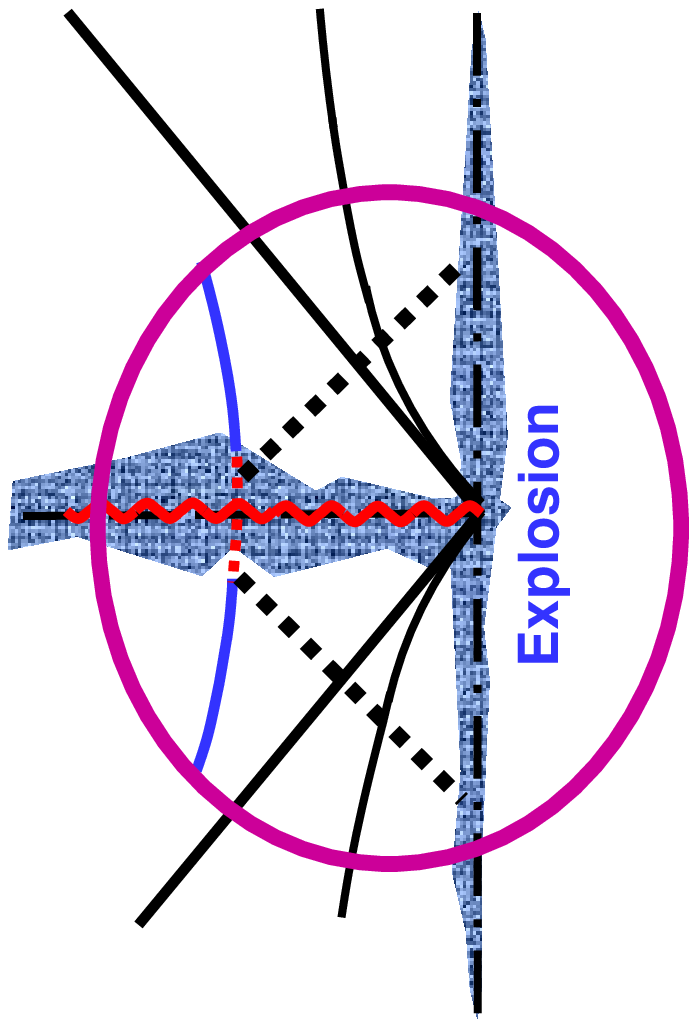}}
\hspace{-1.25in}\scalebox{.30} {\includegraphics{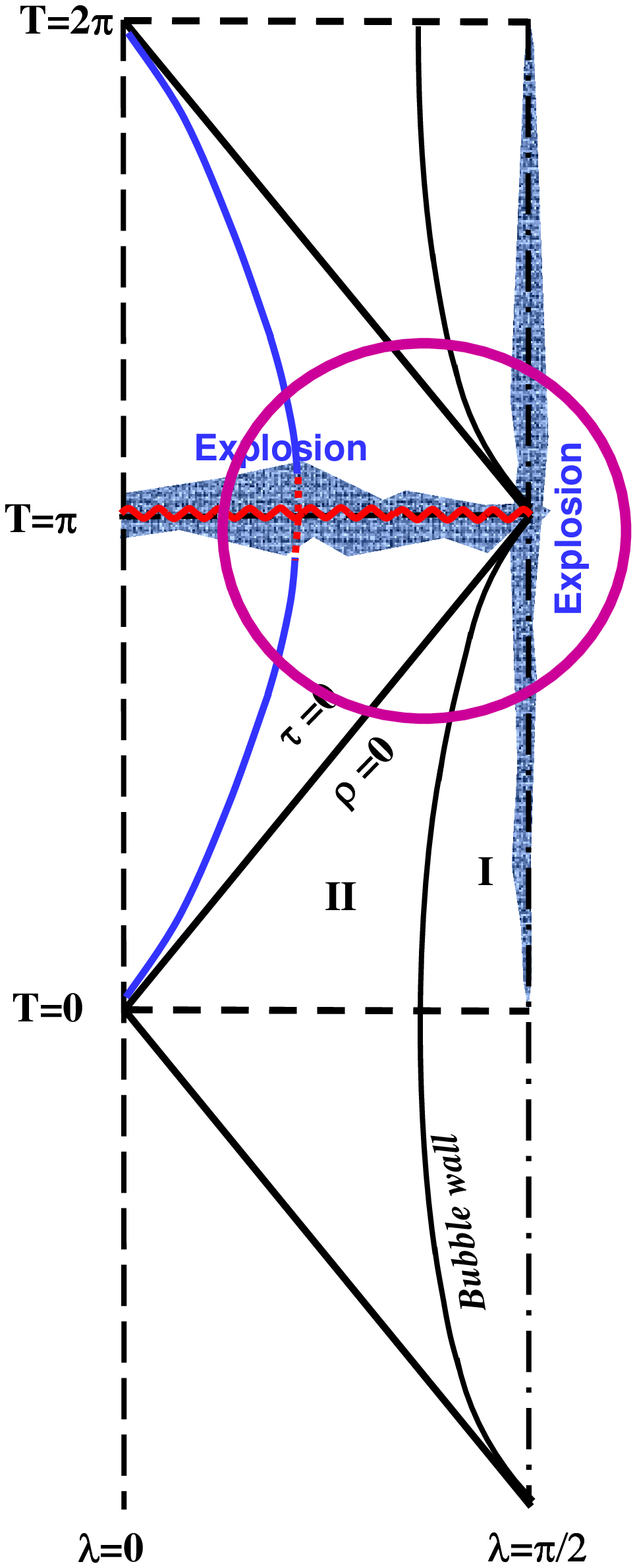}}
\hspace{-1.05in}\scalebox{.35} {\includegraphics{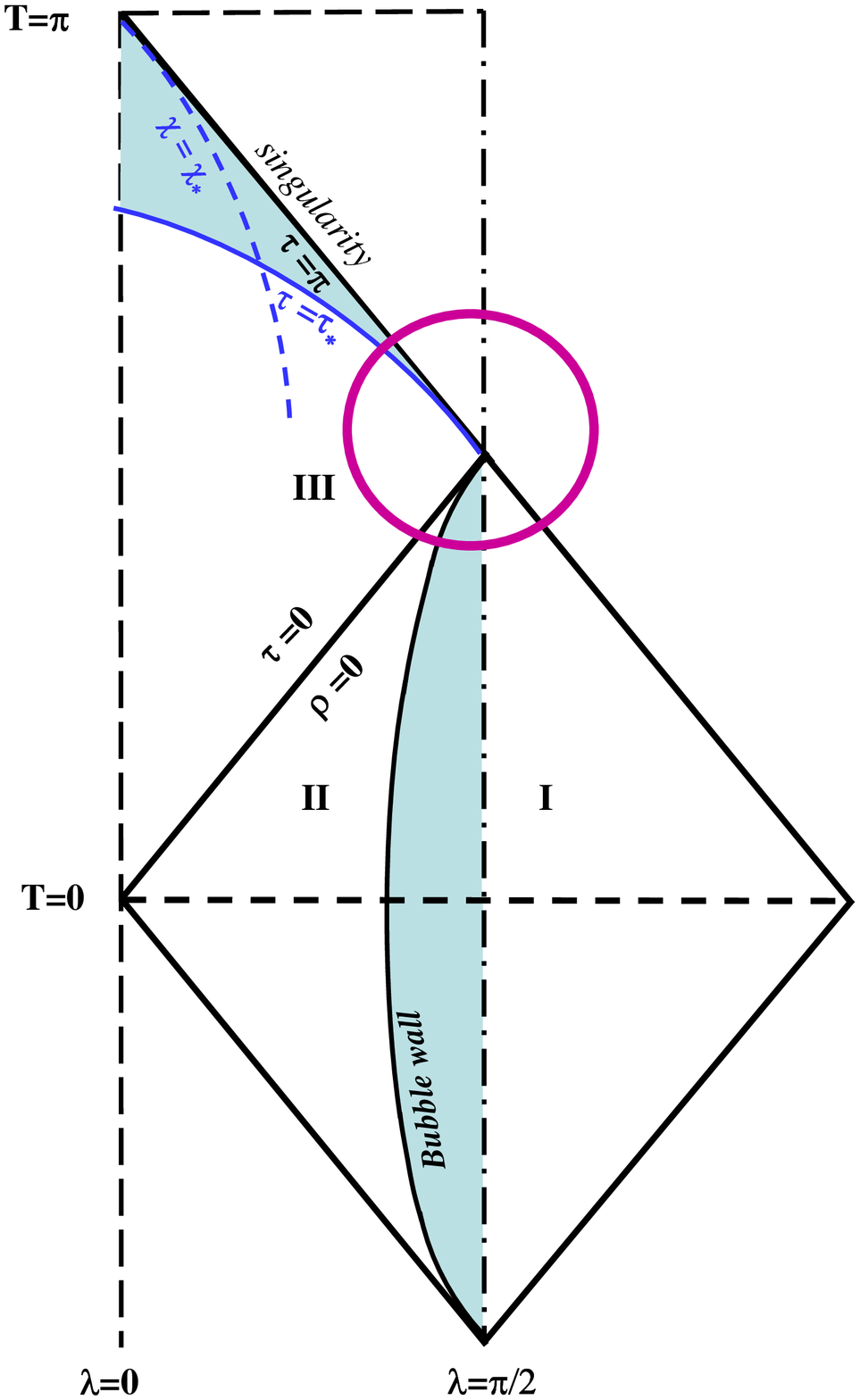}}
\caption{Shown on the right panel is a Penrose diagram of the geometry when some matter is added to the FRW universe. In the shaded region, large deviation from the empty FRW are expected. In the middle panel the explosion and its dual are shown. The bulk explosion spills all the way to the AdS boundary. The left panel depicts the view from the dual dS CFT of the formation and evaporation of the BH.}
\end{figure}

The only observables that could survive this epoch are those that have only low frequency components. All the high frequency ones are essentially very similar to the AdS vacuum (analog of Bunch-Davies for the FRW).

In a way, this is saying that, in practice, the boundary theory has limited ability to describe the details of the big crunch big bang transition. It is a tool that allows to think about this in a concrete framework and if the bulk and boundary are truly equivalent also shows that the transition can occur in a unitary way. However, the amount of information that is passed on is small, similar to the situation in BH evaporation. If you write a message and send it to a BH, it will come out but to read it in the Hawking radiation will be practically impossible. Something similar is true for the crunching FRW.

\section{Summary and Conclusions}

We have proposed a scenario for describing a quantum big crunch to big bang transition. In this scenario, the resolution of the would be classical singularity is due to quantum effects that get strong and cause the semiclassical description to break down.

While we cannot describe the detailed evolution through the transition microscopically we can describe it in an effective way by using thermodynamic potentials. We use entropy bounds to determine the onset of the transition; when they are saturated, the universe, viewed as a quantum state, ceases to be semiclassically (meta)stable. The more stable phase is a collection of relativistic particles, or, in other words, radiation, that occupy the same region of spacetime.

This leads us to propose that an explosion connects the contracting phase to the expanding one. Here, an explosion means fast conversion of the energy in a certain region of spacetime into radiation. After the radiation has been created, the semiclassical description becomes valid again, the geometry responds to the radiation in the expected way, the universe expands and the transition is completed.

We discuss our scenario in three examples: collapsing dust, a brane universe falling into a bulk black hole in anti-de Sitter space, and a contracting universe filled with a negative cosmological constant and a small amount of matter. We briefly discuss the late time observables  that may carry some information about the state of the universe before the transition.

\section*{Acknowledgements}
We thank Jose Barbon, Juan Maldacena, Eliezer Rabinovici, Simon Ross and Joan Simon for discussions. The research of RB was supported by the Israel Science Foundation grant no. 239/10. MSS is supported by a EURYI award of the European Science Foundation. We thank KITP, University of California, Santa Barbara, for hospitality and support. This research was supported in part by the National Science Foundation under Grant No. NSF PHY11-25915.


\begin{thebibliography}{99}

\bibitem{rev1}
R.~C.~Tolman, ``Relativity, Thermodynamics, and Cosmology," Oxford: Clarendon Press, 1934.

\bibitem{rev2}
  E.~M.~Lifshitz and I.~M.~Khalatnikov,
  ``Investigations in relativistic cosmology,''  Adv.\ Phys.\  {\bf 12}, 185 (1963).  

\bibitem{rev3}
  V.~A.~Belinsky, I.~M.~Khalatnikov and E.~M.~Lifshitz,
  ``Oscillatory approach to a singular point in the relativistic cosmology,''  Adv.\ Phys.\  {\bf 19}, 525 (1970).  

\bibitem{rev4}
  S.~W.~Hawking and G.~F.~R.~Ellis,
  ``The Large scale structure of space-time,''  Cambridge University Press, Cambridge, 1973

\bibitem{pbb}
  M.~Gasperini and G.~Veneziano,
  ``The Pre - big bang scenario in string cosmology,''  Phys.\ Rept.\  {\bf 373}, 1 (2003)  [hep-th/0207130].  


\bibitem{ekpyrotic}
  J.~-L.~Lehners,
  ``Ekpyrotic and Cyclic Cosmology,''  Phys.\ Rept.\  {\bf 465}, 223 (2008)  [arXiv:0806.1245 [astro-ph]].  

\bibitem{timelike}
  M.~Berkooz and D.~Reichmann,
  ``A Short Review of Time Dependent Solutions and Space-like Singularities in String Theory,''  Nucl.\ Phys.\ Proc.\ Suppl.\  {\bf 171}, 69 (2007)  [arXiv:0705.2146 [hep-th]].  

\bibitem{Hertog1}
  T.~Hertog and G.~T.~Horowitz,
  ``Towards a big crunch dual,''  JHEP {\bf 0407}, 073 (2004)  [hep-th/0406134].  


\bibitem{Hertog2}
  T.~Hertog and G.~T.~Horowitz,
  ``Holographic description of AdS cosmologies,''  JHEP {\bf 0504}, 005 (2005)  [hep-th/0503071].  


\bibitem{Hertog3}
  B.~Craps, T.~Hertog and N.~Turok,
  ``On the Quantum Resolution of Cosmological Singularities using AdS/CFT,''  arXiv:0712.4180 [hep-th].  

\bibitem{Rabbarb1}
  J.~L.~F.~Barbon and E.~Rabinovici,
  ``Holography of AdS vacuum bubbles,''  JHEP {\bf 1004}, 123 (2010)  [arXiv:1003.4966 [hep-th]].  


\bibitem{Maldacena}
  J.~Maldacena,
  ``Vacuum decay into Anti de Sitter space,''  arXiv:1012.0274 [hep-th].  

\bibitem{Harlow}
  D.~Harlow and L.~Susskind,
  ``Crunches, Hats, and a Conjecture,''  arXiv:1012.5302 [hep-th].  

\bibitem{Rabbarb2}
  J.~L.~F.~Barbon and E.~Rabinovici,
``AdS Crunches, CFT Falls And Cosmological Complementarity,''  JHEP {\bf 1104}, 044 (2011)  [arXiv:1102.3015 [hep-th]].  


\bibitem{Rabbarb3}
  R.~Auzzi, S.~Elitzur, S.~B.~Gudnason and E.~Rabinovici,
  ``Time-dependent stabilization in AdS/CFT,''  JHEP {\bf 1208}, 035 (2012)  [arXiv:1206.2902 [hep-th]].  


\bibitem{ramyrev}
  R.~Brustein,
  ``Cosmological entropy bounds,''  Lect.\ Notes Phys.\  {\bf 737}, 619 (2008)  [hep-th/0702108].  


\bibitem{hol1}
  J.~M.~Maldacena,
  ``The Large N limit of superconformal field theories and supergravity,''  Adv.\ Theor.\ Math.\ Phys.\  {\bf 2}, 231 (1998)  [hep-th/9711200].

\bibitem{hol2}
  O.~Aharony, S.~S.~Gubser, J.~M.~Maldacena, H.~Ooguri and Y.~Oz,
  ``Large N field theories, string theory and gravity,''  Phys.\ Rept.\  {\bf 323}, 183 (2000)  [hep-th/9905111].  


\bibitem{hol3}
  E.~Witten,
  ``Anti-de Sitter space and holography,''  Adv.\ Theor.\ Math.\ Phys.\  {\bf 2}, 253 (1998)  [hep-th/9802150].

\bibitem{Bek1}
  J.~D.~Bekenstein,
  ``A Universal Upper Bound on the Entropy to Energy Ratio for Bounded Systems,''  Phys.\ Rev.\ D {\bf 23}, 287 (1981).  

\bibitem{Boussorev}
  R.~Bousso,
  ``The holographic principle,''
  Rev.\ Mod.\ Phys.\  {\bf 74}, 825 (2002)
  [arXiv:hep-th/0203101].


\bibitem{Bek2}
  J.~D.~Bekenstein,
  ``Is The Cosmological Singularity Thermodynamically Possible?,''  Int.\ J.\ Theor.\ Phys.\  {\bf 28}, 967 (1989).  

\bibitem{Causal}
  R.~Brustein and G.~Veneziano,
  ``A Causal entropy bound,''  Phys.\ Rev.\ Lett.\  {\bf 84}, 5695 (2000)  [hep-th/9912055].  




\bibitem{species1}
  T.~Jacobson,
  ``Black hole entropy and induced gravity,''  gr-qc/9404039.  
\bibitem{species2}
  J.~D.~Bekenstein,
  ``Do we understand black hole entropy?,''  gr-qc/9409015.  
\bibitem{species3}
  G.~Dvali,
  ``Black Holes and Large N Species Solution to the Hierarchy Problem,''  Fortsch.\ Phys.\  {\bf 58}, 528 (2010)  [arXiv:0706.2050 [hep-th]].  

\bibitem{dvaliredi}
  G.~Dvali and M.~Redi,
  ``Black Hole Bound on the Number of Species and Quantum Gravity at LHC,''  Phys.\ Rev.\ D {\bf 77}, 045027 (2008)  [arXiv:0710.4344 [hep-th]].  


\bibitem{BDV}
  R.~Brustein, G.~Dvali and G.~Veneziano,
  ``A Bound on the effective gravitational coupling from semiclassical black holes,''  JHEP {\bf 0910}, 085 (2009)  [arXiv:0907.5516 [hep-th]].  

\bibitem{correspondence1}
  G.~T.~Horowitz and J.~Polchinski,
  ``A Correspondence principle for black holes and strings,''  Phys.\ Rev.\ D {\bf 55}, 6189 (1997)  [hep-th/9612146].  

\bibitem{correspondence2}
  T.~Damour and G.~Veneziano,
  ``Selfgravitating fundamental strings and black holes,''  Nucl.\ Phys.\ B {\bf 568}, 93 (2000)  [hep-th/9907030].  

\bibitem{Gibbons}
  G.~W.~Gibbons and S.~W.~Hawking,
  ``Cosmological Event Horizons, Thermodynamics, And Particle Creation,''
  Phys.\ Rev.\  D {\bf 15}, 2738 (1977).


\bibitem{Randall:1999vf}
L.~Randall and R.~Sundrum, ``An alternative to compactification,'' Phys.\ Rev.\
Lett.\  {\bf 83}, 4690 (1999) [arXiv:hep-th/9906064].

\bibitem{mirage1}
P.~Kraus, ``Dynamics of anti-de Sitter domain walls,'' JHEP {\bf 9912}, 011 (1999)
[arXiv:hep-th/9910149].

\bibitem{mirage2}
A.~Kehagias and E.~Kiritsis, ``Mirage cosmology,'' JHEP {\bf 9911}, 022 (1999)
[arXiv:hep-th/9910174].



\bibitem{Witten}
E.~Witten,
``Anti-de Sitter space, thermal phase transition, and confinement in gauge
theories,'' Adv.\ Theor.\ Math.\ Phys.\ {\bf 2} (1998) 505.



\bibitem{tmax}
 R.~Brustein, D.~Eichler and S.~Foffa,
  ``A braneworld puzzle about entropy bounds and a maximal temperature,''
  Phys.\ Rev.\ D {\bf 71}, 124015 (2005)
  [arXiv:hep-th/0404230].











\end{thebibliography}
\end{document}